# GENMR: Generalized Query Processing through Map Reduce In Cloud Database Management System

Shweta Malhotra, Mohammad Najmud Doja, Bashir Alam, Mansaf Alam

Jamia Millia Islamia, New Delhi, India

Shweta.mongia@yahoo.com, mndoja@gmail.com, babashiralam@gmail.com, Mansaf_alam2002@yahoo.com

**Abstract:** Big Data, Cloud computing, Cloud Database Management techniques, Data Science and many more are the fantasizing words which are the future of IT industry. For all the new techniques one common thing is that they deal with Data, not just Data but the Big Data. Users store their various kinds of data on cloud repositories. Cloud Database Management System deals with such large sets of data. For processing such gigantic amount of data, traditional approaches are not suitable because these approaches are not able to handle such size of data. To handle these, various solutions have been developed such as Hadoop, Map Reduce Programming codes, HIVE, PIG etc. Map Reduce codes provides both scalability and reliability. But till date, users are habitual of SQL, Oracle kind of codes for dealing with data and they are not aware of Map Reduce codes. In this paper, a generalized model GENMR has been implemented, which takes queries written in various RDBMS forms like SQL, ORACLE, DB2, MYSQL and convert into Map Reduce codes. A comparison has been done to evaluate the performance of GENMR with latest techniques like HIVE and PIG and it has been concluded that GENMR shows much better performance as compare to both the techniques. We also introduce an optimization technique for mapper placement problems to enhance the effect of parallelism which improves the performance of such Amalgam approach.

## 1 Introduction

Cloud Database Management System (CDBMS) is one of the probable solutions provided by the IT experts. Many Cloud provider Companies such as Amazon, Yahoo, EMC2, Microsoft, Google, Rackspace etc. provide database services in SQL and NOSQL form. CDBMS is attractive for various reasons as organizations are not bothered about the hardware maintenance, software cost or any administrative cost, they only focus on the efficiency of the business.

Processing of the data on the cloud has become a biggest issue now a days. Traditional database management systems are not able to process such hefty size of data. New technologies such as MapReduce, Hive, PIG, Hadoop etc. are coming out for processing such size of data. But till date, users are very much comfortable with traditional DBMS as they are not aware of the benefits of MapReduce codes.

Map Reduce codes [1] available in the market are attractive due to the benefits like [7], these codes are in the simple Key-value form hence they are easy to use. It is a Cost effective solution for processing large size of data as codes provide parallel processing. Map Reduce codes provide flexibility as it is not based on any schema, data can either be in structured or unstructured form. Another benefit of the Map Reduce codes is they provide scalability as well.

In this paper a generalized model GENMR has been proposed and implemented to give solutions for processing such large sized data for cloud database management system. The key contribution of our work is defined as follows:

1. We propose an efficient approach for processing large amount of data.
2. Layer wise responsibility related to Cloud Database Management System and Architecture of proposed model have also been defined.

3. Our approach can take queries in any of these form SQL, MYSQL, DB2, Oracle.
4. With the help of GENMR compiler queries data converted into MapReduce Key-Value form.
5. This is an efficient way as compare to the other latest techniques such as HIVE and PIG.

The rest of the paper is organized as follows. Section 2, describes the state of the work that has been done so far related to the field of Cloud Database Management System, Big Data. In Section 3, we briefly define our proposed model along with the algorithms which is used for the implementation of GENMR. In section 4, Results and analysis has been described. We analyzed GENMR with latest techniques. Lastly, we conclude our work with future possibilities in section 5.

## 2 Related work

Simple MapReduce [1][2] codes in key-value pair are considered to be a suitable solution for large amount of parallel data processing. An enhancement to the MapReduce codes is being provided with the help of pipelining concept i.e. Whenever Mapper function produces its results in the intermediate form it goes to Reducer function for generating output [3][4] to provide the more parallel processing of data . Authors [5] described the efficient way to process Big data across geographical distributed data centers.

It is difficult to learn the MapReduce codes so systems are provided to convert the queries into Map Reduce form. SQLMR and YSmart[6][7], are the examples of such systems which take queries in SQL form and convert queries into equivalent MapReduce form.

Author in [8] explained one optimization algorithm for cross Rack Optimization for Reducer program. Here, generalized model considers Mapper function into account as well. Proposed generalized model [11] takes queries in SQL, MYSQL, DB2, Oracle form and converts into MapReduce form with many enhancements. Big data analysis [12] on Cloud has become issue here author provides a solution of Map Reduce algorithm and Bigdata analytic techniques. In Paper [15] author tried to explain the enhancements that going into Cloud Computing world. MySQL provides a way to process and manipulate data but it is not applicable for large amount of data set. In [13] [14], author compared large data set with HIVE[16] and Pig.

## 3 Proposed Model

The problem with the today's world is that users are not aware of Map Reduce kind of codes to process large size of data present at the cloud repositories. Hence, this model comprises of a solution which takes up user queries in RDBMS form like (SQL, DB2, Oracle, MySQL) and with the help of model's compiler module, queries get converted into map reduce form as map reduce is a splendid solution to process large amount of data. As shown in Fig 1, five layer architecture for CDBMS has been proposed [9] [10] in this paper, a detailed working description related to each layer has been provided in Fig 2.

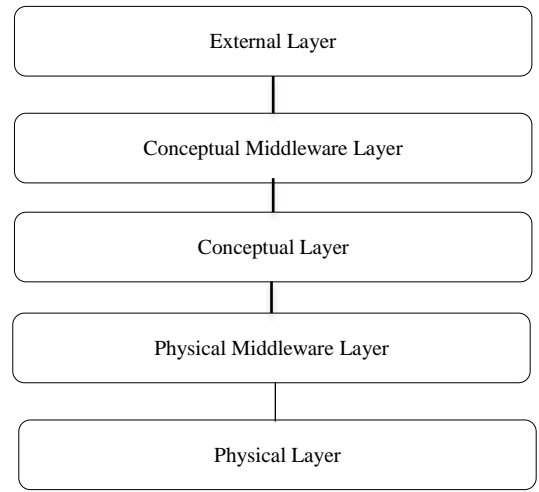

Fig 1. Layers of CDBMS

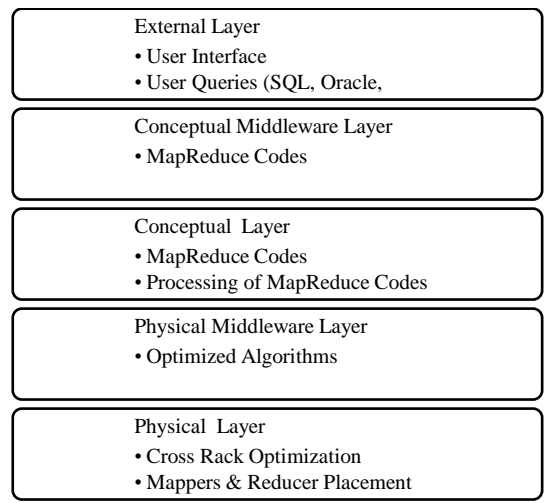

Fig 2. Layer wise responsibility

## 3.1 External Layer: User interface

External Layer is the only layer which is closest to the user and provide interfacing. The main function of this Layer is to provide the transparency and to manage different types of users. User sends their queries in the form of SQL, DB2, Oracle, MySQL where data is already stored into the system. Existed data is pre-partitioned horizontally and stored into number of Data nodes of the Racks to have parallel and distributed processing as explained by algorithm 1. Algorithm 1 also described the way data is stored in Inter-Rack or Intra-Rack to have Inter or Intra Rack Communication. Table 1 & Table 2 comprised of a symbols and assumptions used throughout the paper.

Table 1. Symbol used

| Sr. No. | Symbol used | Definition/explanation |
|---|---|---|
| 1 | $Rack_1, Rack_2, Rack_3, \ldots Rack_n$ | n number of Racks. |
| 2 | $d_{11}, d_{12}, \ldots d_{1m}$ | Each Rack consist of m no of Datanodes, example shows these Datanodes are of $Rack_1$ |
| 3 | $Data_1, Data_2$ | Data present on the Datanodes |
| 4 | M | Mapper function |
| 5 | R | Reducer Function |

Table 2: Assumptions used

| Sr. No. | Assumptions |
|---|---|
| 1 | There is only one client's data present on the Cloud Database. |
| 2 | Client data partitioned on the FCFS basis. |
| 3 | One Data row is present at one Datanode of a Rack |
| 4. | Datanode capacity is q …. q rows can be kept on that Datanode. |

### Algorithm 1. Pre-partitioning

**Input:** Data is stored horizontally in rows. One row is stored at one DataNode.
Data is placed as per the Datanode capacity
Datanode capacity is q rows.
**Output:** Partition the data on the Intra racks i.e. Users data is placed at the Datanodes of same Racks
$d_{1\_rack1\_row1}, d_{1\_rack1\_row2}, \ldots\ldots\ldots\ldots\ldots d_{1\_rack1\_rowq}$,
$d_{2\_rack1\_rowq+1}, d_{2\_rack1\_rowq+2}, \ldots\ldots\ldots d_{2\_rack1\_rowq+q}$,
$d_{z/q\_rack1\_row(z-i-2)}, d_{z/q\_rack1\_row(z-i-1)}, \ldots\ldots\ldots d_{z/q\_rack1\_rowz}$,
or Inter Rack i.e. Users data is placed at the Datanodes of different Racks to have parallel processing.
$d_{1\_rack1\_row1}, \ldots d_{i\_rack1\_rowq}, \ldots d_{i+1\_rack2\_rowq+1}$,
$\ldots d_{i+i\_rack2\_rowq+q}, \ldots d_{z/q\_rackn\_rowz}$,

1. **Procedure : Pre-Partitioning**
2. For user's data
3. Case 1: *Intra rack*
4. If total Data size is z.
5. Total number of Datanodes required on that particular rack will be
   Total Datanodes = z/q…………………………..(i)
6. Until all the data is placed at the Datanodes of the Rack.
7. Case 2: *Inter Rack.*
8. Total number of Datanodes required on that all the Racks will be same i.e. Total Datanodes = z/q
9. for i=1to n … for n number of Racks
10. for j= 1to m ….. for m datanodes
11. Data is partitioned as to have total datanodes = z/q
12. Until all the data is placed at the Datanodes of the Racks.
13. End of For loop
14. End of For Loop

## 3.2 Conceptual Middleware Layer: Any Database to Map Reduce Compiler

This layer provides interoperability means it hides the availability of different databases to the users and operates irrespective of the underlying databases available. User's process their queries in the Databases languages in which they are comfortable. Users till date are comfortable with RDBMS tools but RDBMS is not a probable solution for processing large amount of Data. Users are not aware of new technologies like Map Reduce Programming Paradigm, Hive, Pig, HBase which can process large amount of data. This layer provides the facility to the users such that their queries are converted into NOSQL Map-Reduce key-value form. Compiler takes input queries from the User interface which is at the external layer. It converts that query into map reduce codes. Query takes pre-partitioned data from the text file stored at the Datanodes of the Racks etc. On the basis of queries again partitioning is applied. Table 3 has the detail of queries considered in this work and the corresponding Key-value pairs defined by the Model's compiler.

## 3.3 Conceptual Layer: Data Processing

This layer deals with actual processing of data. At this layer actual processing of key value pair is being done. Reducer will be applied to the partitioned intermediate data. Table 3 comprised of the queries for data processing and with the help of conceptual Middleware Layer's Compiler these queries converted into key value pair. Now, at conceptual Layer reducer program takes the key-value pair and give results accordingly as given by the algorithm 2.

Table 3: Database Queries

| Sr. No. | Query | Map Reduce- Key-Value Pair | |
|---|---|---|---|
| 1 | Select * from Table Name where **Column name**= value | Key= **Column name** | Value= all other fields name except key column name |
| 2 | Select Count **Column name** from Table Name | Key=**Column name** | Value=1 |
| 3 | Select Distinct **Column name** from Table Name | Key=**Column name** | Value=1 |
| 4 | Select Upper **Column name** from Table Name | Key=**Column name** | Value=1 |
| 5 | Select substring **Column name** from Table Name | Key=**Column name** | Value=1 |
| 6 | Select Count **Column Name** from Table Name where Column Name = value | Key=**Column name** | Value=1 |
| 7 | Select Distinct **Column Name** from Table Name + where Column Name = value | Key=**Column name** | Value=1 |
| 8 | Select Upper **Column Name** from Table Name + where Column Name = value | Key=**Column name** | Value=1 |
| 9 | Select substring **Column Name** from Table Name + where Column Name = value | Key=**Column name** | Value=1 |
| 10,11,12,13 | Select +( Count/ Distinct/ Upper/ substring)+**Column Name** from Table Name+ where Column Name = value +(and) + Column Name= Value | Key= **Column name** | Value=1 |
| 14,15,16,17 | Select +( Count/ Distinct/ Upper/ substring)+**Column Name** from Table Name+ where Column Name = value + (or) + Column Name= Value | Key=**Column name** | Value=1 |
| 18 | Select * from Table Name Orderby **Column Name** Asc/ Desc | Key=**Column name** | Value= all other fields name except key column name |
| 19 | Groupby | Key=**Column name** | Value=1 |

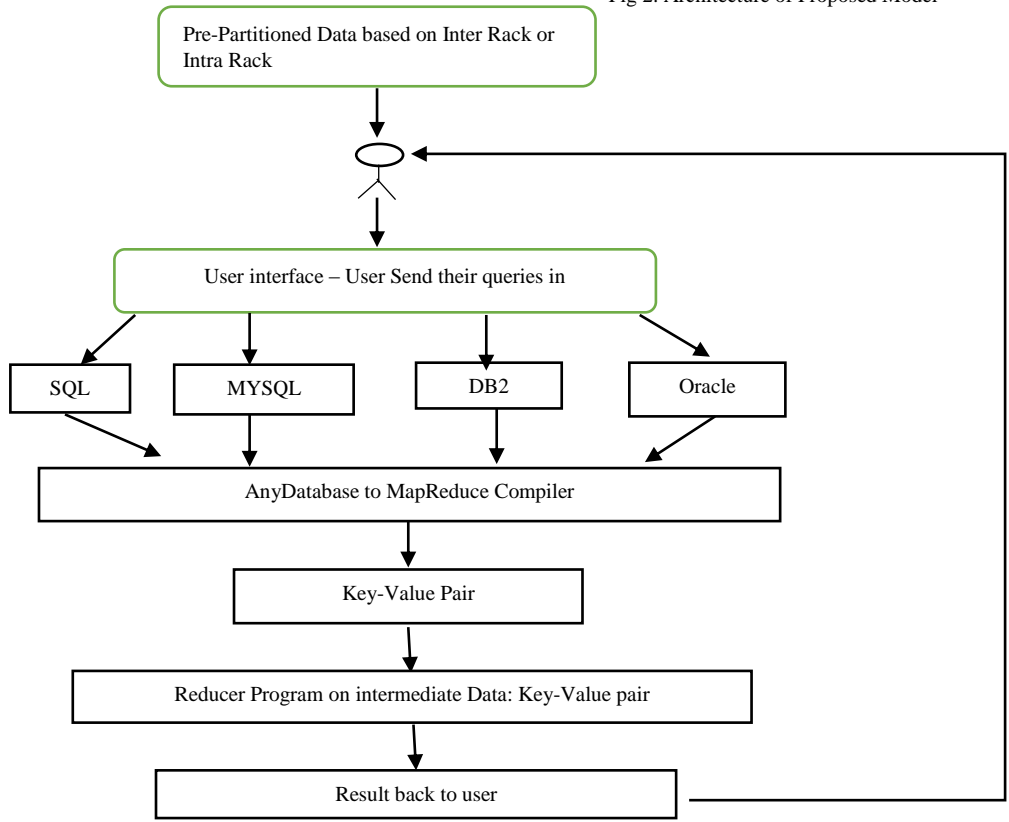

Fig 2. Architecture of Proposed Model

**Algorithm 2 – <u>Generation of Key Value Pair: Any Database to Map Reduce compiler</u>**

Queries are applied on the pre-partitioned data. Proposed model can take queries syntax of SQL, MYSQL, ORACLE and DB2.

Input: Queries in SQL, DB2, MYSQL, Oracle
Output: Key-Value Pair & Processed result back to user.
For Queries in SQL, DB2, MYSQL, ORACLE, following algorithm explained the generation of the Key-Value Pair. Mapper program is used to generate key_value pair and Reducer program gives results back to the user on the basis of key_value pair.

1. **Procedure: <u>Any Database to MapReduce compiler</u>**
2. For each query, Mappers will generate the Key-value pair
3. If Query == "*SELECT * FROM Table_Name WHERE Column_Name == Value*"
       Key = Column Name
       Value == all the fields of Table except key Column name
       Result = Key + value
4. ElseIf Query == "*SELECT COUNT Column_Name FROM Table_Name + WHERE Column_ name1==Value + AND/ OR + Column_Name2==Value*"
       Key= Column_Name
       Value= 1
       Result= Sum (values)
5. ElseIf Query == "*SELECT DISTINCT Column_Name FROM Table_Name + WHERE Column_ name1==Value + AND/ OR + Column_Name2==Value*"
       Key= Column_Name
       Value= 1
       Result=Sum (Distinct value of key Column Name)
6. ElseIf Query == "*SELECT UPPER Column_Name FROM Table_Name + WHERE Column_ name1==Value + AND/ OR + Column_Name2==Value*"
       Key= Column_Name
       Value= 1
       Result= Upper (Column_Name)
7. ElseIf Query == "*SELECT SUBSTRING Column_Name FROM Table_Name + WHERE Column_ name1==Value + AND/ OR + Column_Name2==Value*"
       Key= Column_Name
       Value= 1
       Result= Substring (Column_Name)
8. Elseif Query== "*SELECT Column_Name == Value ORDERBY Asc/ Desc*"
       Key= column name
       Value= all the fields of Table except key Column name
       Result = Asc/ Desc( Column_Name)
9. End of nested if-else
10. End of For loop for queries.

**3.4 Physical Middleware layer and Physical Layer: Mapper Reducer Placement and Storage issues**:

Two important aspects related to storage like Inter Rack or Intra Rack communication and Mapper and Reducer function Placement problems are considered at these layers. Physical Middleware layer provides the interoperability but main storage related issues are dealt on Physical Layer.

Initially, at the physical layer data is being pre-partitioned as per the algorithm 1.

| Rack1 | Rack2 | Rack3 |
|---|---|---|
| • Datanode10 | • Datanode20 | • Datanode30 |
| • Datanode11 | • Datanode21 | • Datanode31 |
| • Datanode12 | • Datanode22 | • Datanode32 |
| • Datanode13 | • Datanode23 | • Datanode33 |
| • Datanode16 | • Datanode24 | • Datanode34 |
| • Datanode17 | • Datanode25 | • Datanode35 |
| • Datanode18 | • Datanode26 | • Datanode36 |
| • Datanode19 | • Datanode27 | • Datanode37 |
|  | • Datanode28 | • Datanode38 |
|  | • Datanode29 | • Datanode39 |

Fig 3. Inter Rack and Intra Rack Communication

Now, for each rack the amount of data sent from or coming into one rack to another is defined based on equation 1, these functions are defined in terms of mapper and reducer.

$$f_i(m_i, r_i) = m_i d_i r_i \quad \text{-- eq(1)}$$

Where m represents Mapper function, r represents Reducer Function, d represents the Datanode,

To place Mappers and reducer function three possibilities can be considered.

1. Mapper and reducer is placed at the same DataNode of same Rack where Users Data is present.

2. Mapper is placed at the same DataNode where data is present but reducer is placed at the other DataNode of same rack.
3. If mapper is placed at the same DataNode of the same rack where data is present but reducer is placed at the different DataNode of the different rack.

Where i=1, 2, 3………………., n are the number of racks.

Cross rack optimization [5] considered only Reducer Placement Problems. In this work both Mapper and Reducer Placement Problems on the DataNodes of the Racks have been considered. Earlier Reducer function problem on the Racks DataNodes were not been considered.

**Algorithm 3: Mapper Function Placement Problem**
Input: Data Size and Datanode capacity.
Output: Mappers are placed on the Datanodes for data Processing.

1. Procedure: Mapper Function Placement Problem
2. If Datanode capacity is q …. q Rows can be kept on that data node.
3. If total data size is z.
4. Then total number of datanodes required on that particular rack will be:
    Total Datanodes = z/q
5. No. of mappers= total datanodes for that data
6. End if

**Algorithm 4: Reducer Function Placement Problem**
Input: Mappers function gives output in the form of Key value pair i.e. intermediate data.
Output: Reducer Functions are placed at the intermediate data.

1. Procedure: Reducer Function Placement Problem
2. For each Query as per the Algorithm 2 Map Reduce compiler will produce some intermediate data in the form of Key-value pair.
3. As per the query some aggregation function is applied on the intermediate data.
4. End of For loop

Reducer function gives back Result of the aggregation function to the user.

4. **Results and Performance Analysis**

Firstly, we analyzed the GENMR with latest technologies for processing a huge amount of data such as HIVE and PIG and in the other section GENMR's processing power is being evaluated on the basis of inter or intra Rack Communication and based on the Mappers function Placement Problem to have parallel processing.

*1. Comparison with HIVE and PIG*

HIVE and PIG are being considered as a suitable choice for such situations. HIVE[16] is considered as SQL-Like language in which users put their queries in SQL form and with the help of Hadoop framework their queries internally gets converted into MapReduce and users without knowing this fact will gets results out of their queries. Similarly in PIG, PIG is a scripting type Language where users again gives their queries in the scripts form and these queries again with the help of Hadoop framework internally gets converted into MapReduce form and users gets their results back.

Here, in this Implemented model, we implemented GENMR in C#, with the help of .NET framework. The advantage with this system is that users can give their queries in any Database forms (MYSQL, SQL, DB2, Oracle). We have incorporated the syntax of these Relational Databases.

In order to evaluate the performance of GENMR system with PIG and HIVE Table4 shows the system requirements that is being considered in this work.

Table 4. System Requirement

| System | Requirement |
|---|---|
| GENMR | Windows 8, .NET framework with C#, one database table with 325 Rows. |
| HIVE | Pseudo-Hadoop cluster with HIVE-0.13.1 version, one database table with 325 Rows. |
| PIG | Pseudo-Hadoop cluster with HIVE-0.12.0 version, one database table with 325 Rows. |

Table 5 describes the queries which is being taken for comparison between GENMR, HIVE and PIG. Below is the summary of conclusion:
1. For Select * Query it has been observed that HIVE takes approximate double time as compared to GENMR.
2. For other queries PIG performs very poorly as compare to HIVE. For COUNT, with and without WHERE clause it has been observed

that GENMR performs approximately four times better then HIVE.
3. For other queries also it has been observed that GENMR outperforms then both the languages.

Table 5. Queries considered for comparison between GENMR, HIVE, PIG.

| Sr. No. | Query: | GENMR (Time in sec) | PIG (Time in sec) | HIVE (Time in sec) |
|---|---|---|---|---|
| 1 | Select * from Teachers where state is Andhra Pradesh | 16.55 | 29 | 38.05 |
| 2 | Select Count state from teachers | 7.92 | 41 | 32.81 |
| 3 | Select distinct state from teachers | 19.35 | 56 | 33.14 |
| 4 | SELECT UPPER(State) FROM teachers | 13.58 | 46 | 22.87 |
| 5 | SELECT SUBSTRING(State,1,5) FROM teachers | 8.00 | 43 | 22.28 |
| 6 | SELECT COUNT(State) FROM teachers WHERE State = 'Andhra Pradesh' | 5.60 | 56 | 33.40 |
| 7 | SELECT DISTINCT(State) FROM teachers WHERE State = 'Andhra Pradesh' | 20.28 | 42 | 30.50 |
| 8 | SELECT COUNT(State) FROM teachers WHERE State = 'Andhra Pradesh ' AND School_Type='Secondary School' | 3.52 | 42 | 31.37 |
| 9 | SELECT COUNT(State) FROM teachers WHERE State = 'Andhra Pradesh ' OR School_Type='Secondary School' | 1.88 | 42 | 30.64 |
| 10 | SELECT * FROM teachers ORDER BY State ASC | 3.33 | 122 | 29.92 |

In general, it has been observed that GENMR model gives better results as compare to the latest techniques Pig and Hive. Fig 5 shows the graph of Data processing time taken for the 10 queries given in table 5. For all the queries the proposed model GENMR has been considered a better solution.

*2. Inter Rack and Intra Rack communication*

For inter Rack Communication and Intra Rack Communication as per the general observation it has been observed that intra Rack communication takes less time comparatively to inter Rack communication. Further, an analysis has been done to see the effect of parallelism. Mapper functions placed on the Data provides the parallelism for data processing.

Mapper functions defined depends upon the Datanode capacity and Partitioning of data depends upon the Datanode size. For example, If a Rack is having capacity of 1 TB =1000 GB. Let the capacity of one Datanode at the rack is 100 GB, If 1 TB data is divided it will take 10 datanodes. So, it will require 10 Mappers and let the capacity of one Datanode at the Rack is 200 GB. If 1 TB data is divided it will take 5 datanodes. So it will require 5 Mappers in this case.

Three sets have been taken for observing the effect of data parallelism.
For 325 rows of a table 6 shows the scenarios that we have considered to see the effect of parallelism.

Table 6. Scenarios for observing Parallelism

| Datanode Capacity | 10 Rows | 30 Rows | 50 Rows |
|---|---|---|---|
| Mappers Required | 32 | 10 | 6 |

It has been concluded that GENMR performs better with 10 mappers as compare to the 32 and 6 mappers, for processing data on many queries.
Table 7 are the list of queries that we considered in this work. Below is the summary of conclusion:
1. For Intra Rack communication GENMR always performs better as compare to Inter Rack Communication.
2. For queries like COUNT, DISTINCT, UPPER, SUNSTRING, ORDERBY with and without WHERE Clause GENMR with 10 Mappers outperforms then 32 Mappers and 6 Mappers.
3. For Queries Select *, COUNT+ WHERE+ AND/OR it has been observed that System with 32 Mappers gives better results.

In general GENMR performs better with 10 Mappers. For 32 Mappers, it takes more communication overhead so it takes more time than system with 10 mappers and 6 Mappers. Secondly System with 6 Mappers shows more time because it is showing less parallelism hence more time is taken for data processing for 6 Mappers as compare to 10 Mappers.

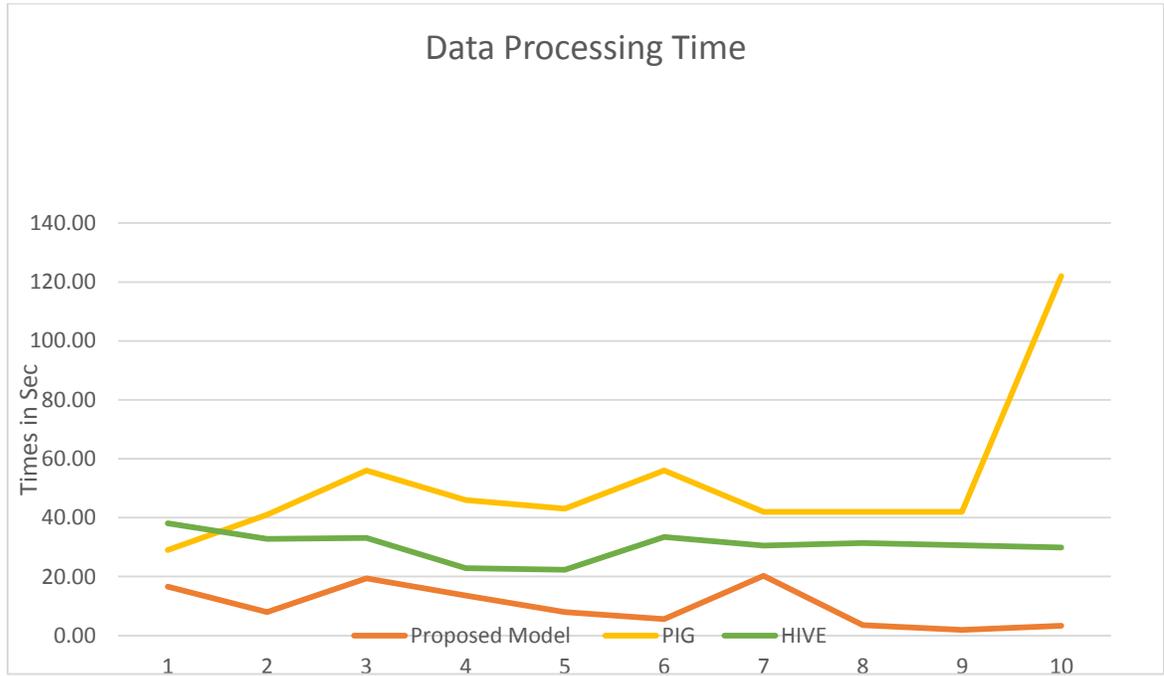

**Fig 4. Data Processing time of Proposed Model, Pig Language and Hive Language**

Table 7. Queries Considered for Inter Rack and Intra Rack Communication and for Parallel Processing

| Sr. No. | QUERIES | Intra | inter | intra | inter | intra | inter |
|---|---|---|---|---|---|---|---|
| | | 32 Mappers (Processing Time in Sec) | | 10 Mappers (Processing Time in Sec) | | 6 Mappers ((Processing Time in Sec) | |
| 1 | SELECT * FROM teachers WHERE State='Andhra Pradesh' | 16.55 | 36.33 | 26.72 | 15.95 | 20.57 | 23.30 |
| 2 | SELECT COUNT(State) FROM teachers | 7.92 | 18.60 | 5.30 | 15.55 | 29.22 | 20.88 |
| 3 | SELECT DISTINCT(State) FROM teachers | 19.35 | 16.00 | 15.47 | 26.40 | 13.50 | 17.32 |
| 4 | SELECT UPPER(State) FROM teachers | 13.58 | 26.15 | 4.10 | 16.70 | 18.07 | 21.20 |
| 5 | SELECT SUBSTRING(State,1,5) FROM teachers | 8.00 | 27.10 | 2.70 | 18.72 | 13.53 | 18.32 |
| 6 | SELECT COUNT(State) FROM teachers WHERE State = 'Andhra Pradesh' | 5.60 | 23.63 | 4.52 | 27.82 | 16.07 | 10.55 |
| 7 | SELECT DISTINCT(State) FROM teachers WHERE State = 'Andhra Pradesh' | 20.28 | 34.67 | 20.10 | 38.00 | 19.70 | 37.45 |
| 8 | SELECT UPPER(State) FROM teachers WHERE State = 'Andhra Pradesh' | 4.60 | 14.38 | 31.37 | 27.35 | 11.12 | 27.90 |
| 9 | SELECT SUBSTRING(State,1,4) FROM teachers WHERE State = 'Andhra Pradesh' | 24.25 | 17.82 | 16.60 | 28.67 | 20.08 | 29.33 |
| 10 | SELECT COUNT(State) FROM teachers WHERE State = 'Andhra Pradesh' AND School_Type='Secondary School' | 3.52 | 22.55 | 18.67 | 38.72 | 17.55 | 33.62 |
| 11 | SELECT COUNT(State) FROM teachers WHERE State = 'Andhra Pradesh' OR School_Type='Secondary School' | 1.88 | 29.40 | 12.27 | 30.40 | 17.95 | 31.15 |
| 12 | SELECT * FROM teachers ORDER BY State ASC | 19.53 | 25.10 | 3.33 | 7.47 | 6.15 | 25.75 |

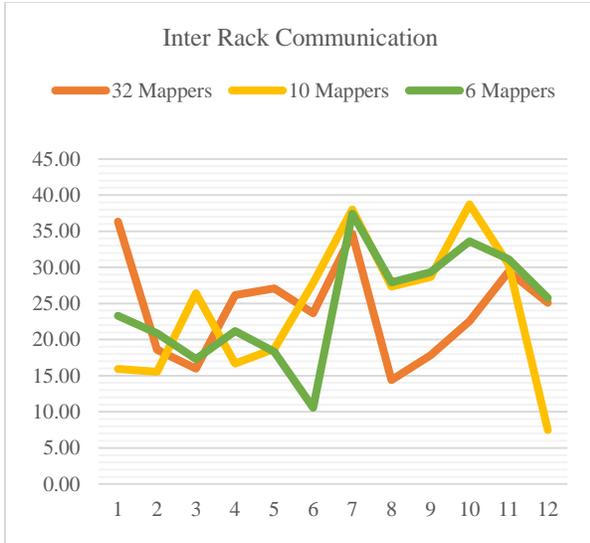

Fig 5: Intra Rack Communication     Fig 6: Inter Rack Communication

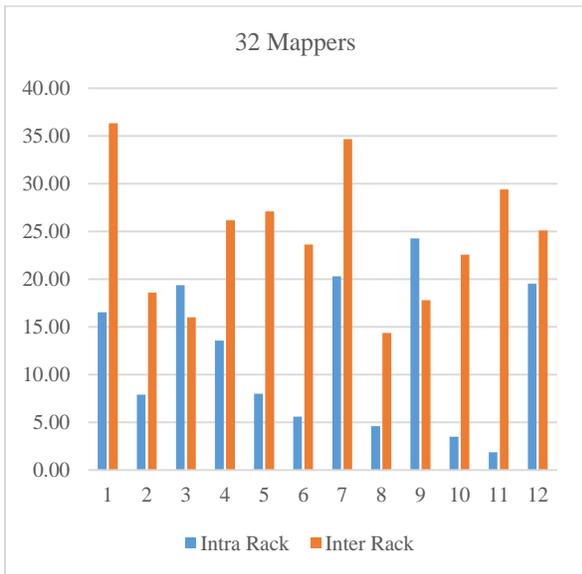 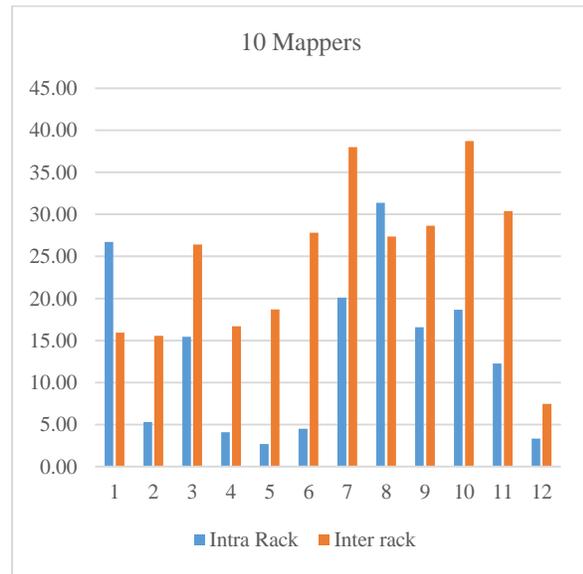

Fig 7: Intra Rack and Inter Rack communication for 32 Mappers     Fig 8: Intra Rack and Inter Rack communication for 10 Mappers

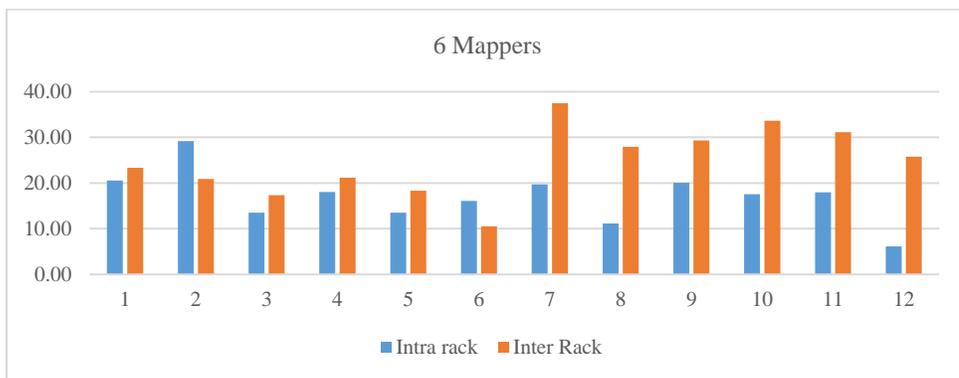

Fig 9: Intra Rack and Inter Rack communication for 6 Mappers

## 5. Conclusion

As users are comfortable with Relational Database languages, in this work a model has been implemented which takes users queries and through the model's compiler these queries gets converted into Map-Reduce key-value form. It is easier to process large amount of data with the help of MapReduce codes as compare to Traditional databases. The model has also been evaluated and compared with the latest technologies in the field of Cloud and Big data i.e Pig and Hive. It has been observed that the GENMR outperforms as compare to the HIVE and PIG. Moreover the GENMR has been analyzed to see the effect of parallelism. So when GENMR system is being used with 10 Mappers it has been observed that system gives much better results as compare to System with 32 Mappers and 6 Mappers.

In Future, as cloud consist of many clients more number of clients can be considered. Compiler can be developed for other complex queries. Instead of linear partitioning hash partitioning can also be applied for client's data.